\documentclass[preprint]{aastex}

\usepackage{epsfig}
\usepackage{amsmath}

\slugcomment{Submitted to ApJ Letters}

\shorttitle{Salmonson}
\shortauthors{Kinematics and GRB980425}

\begin{document}


\title{On the Kinematics of GRB980425 and its association with SN1998bw}


\author{Jay D. Salmonson}
\affil{Lawrence Livermore National Laboratory, Livermore, CA 94550}



\begin{abstract}

In this paper I put forward a model in which GRB980425 is both
associated with SN1998bw and is also a standard canonical (long;
$\sim$ seconds) gamma-ray burst.  Herein it is argued that if
gamma-ray bursts are relativistic jets with the fastest moving
material at the core, then the range of observed jet inclinations to
the line-of-sight produces a range in the observed properties of GRBs,
i.e.~the lag-luminosity relationship.  In particular, if the jet
inclination is high enough, the observed emitter will move slowly
enough to render relativistic beaming ineffective, thus distinguishing
the jet from apparent isotropic emission.  Thus we expect a break in
the lag-luminosity relationship.  I propose that GRB980425 defines
that break.  The position of this break gives important physical
parameters such as the Lorentz factor ($\gamma_{max} \sim 1000$), the
jet opening angle ($\sim 1$ degree), and thus the beaming fraction
($\sim 10^{-4}$).  Estimates of burst rates are consistent with
observation.  If correct, this model is evidence in favor of the
collapsar mode as the progenitor of cosmological, long gamma-ray
bursts.

\end{abstract}


\keywords{gamma rays: bursts --- gamma rays: theory}


\section{Introduction}

GRB980425 and its apparent association with SN 1998bw \citep{gvpk+98}
has drawn much attention among gamma-ray burst researchers.  While the
connection of the gamma-ray burst (GRB) to the supernova (SN) remains
uncertain, it is striking that GRB980425 and SN 1998bw were each, taken
individually, unusual events.

To start, SN 1998bw was a relatively rare and unusually luminous Type I b/c
supernova \citep{gvpk+98}.  It was the brightest radio supernova ever observed
\citep{wl99} which may have been due to relativistic outflow with
Lorentz factor $\gamma \sim 2$ \citep{kfwe+98}.

Furthermore, GRB980425 was an unusual GRB.  It was comprised of a
single, unusually rounded peak \citep{bkhp+98}.  A cool burst, it was
not seen in BATSE's highest energy channel ($> 300$ keV)
\citep{nmb00}.  If this burst is indeed associated with SN1998bw
($z=0.008$), then the burst is apparently vastly weaker than all other
known bursts, with an inferred isotropic gamma-ray energy of $8 \times
10^{47}$ ergs \citep{gvpk+98}. Finally, of particular interest to this
paper, \citet{nmb00} found that the lag of the peak of this burst
between BATSE channels 1 and 3 was exceptionally large: $\Delta
t_{980425} \approx 4.5$ seconds.

In light of the respective idiosyncrasies of these two events, we may
either conclude that we have seen a chance coincidence of two unusual
events, with a probability of $10^{-4}$ or less \citep{gvpk+98}, or
perhaps the discovery of an entirely new type of GRB
\citep{gvpk+98,bkhp+98}.

In this paper I propose a third alternative; that GRB980425/SN1998bw
is a canonical gamma-ray burst, deriving from a relativistic jet
driven by a collapsar \citep{mfw99}, observed at high angle of
inclination.  The idea that this burst was a jet viewed off-axis has
been proposed by several other authors
\citep{ww98,wes99,naka99,hww99}.  However in this paper I show how
GRB980425 may be a canonical gamma-ray burst by its relation to other
bursts on the lag-luminosity relationship discovered by \citet{nmb00}.
Thus this burst need not be a distinct class of GRB or a special case
of a failed collapsar, polluted with excessive baryon entrainment
within the jet \citep{wmf99}.  Physical GRB parameters can be gleaned
from this identification.

\begin{figure}[tb]
\centering
\epsfig{figure=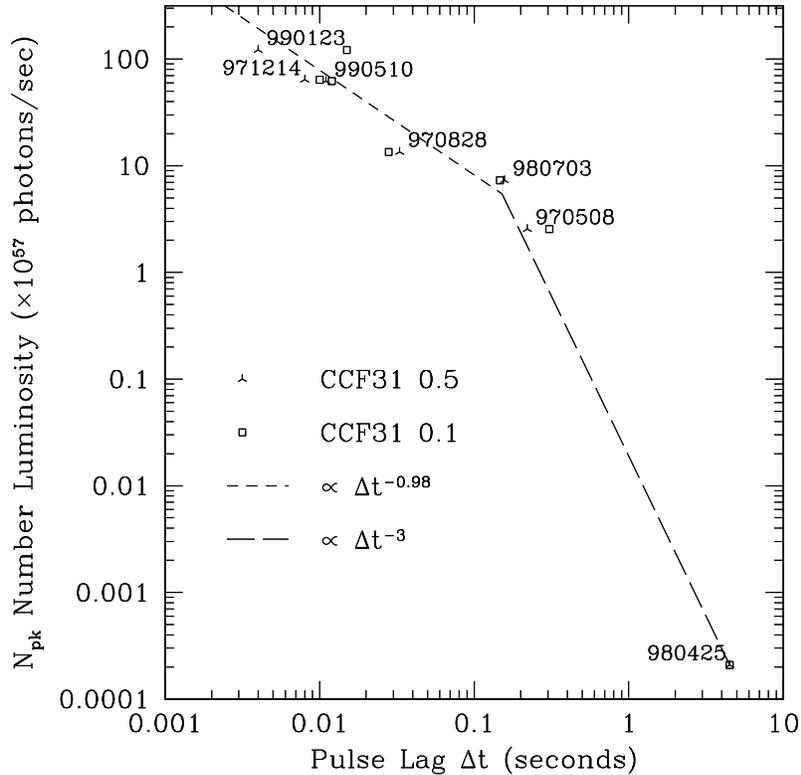, width = 11 cm} % 7 cm }
\caption{Peak photon number luminosity $N_{pk}$ versus spectral pulse
lag for six bursts with known redshifts plus GRB980425.  A break is
inferred by fitting a break slope $\propto \Delta t^{-3}$ to intersect
GRB980425.  Spectral cross-correlation function lags between BATSE
channels 3 and 1 (CCF31) for regions down to 0.5 and 0.1 of peak
intensity were obtained from \citet{nmb00}.  The line of best fit for
0.1 (squares) is $\propto \Delta t^{-0.98}$.  }
\label{lag-lum-980425}
\end{figure}

In brief, if we assume that the core of the jet has the highest
velocity material and that the velocity monotonically decreases with
increasing angle from the core axis, then there will be an angle at
which the $1/\gamma$ aperture imposed by relativistic beaming becomes
comparable to the angular size $\theta_0$ of the emitting region.  At
this angle there will be a change in the observed properties of the
GRB.  In particular, one would expect a break in the lag-luminosity
relationship discovered by \citet{nmb00} and futher interpreted by
\citet{jay00}.  Herein I show that the peak number luminosity's
inverse dependence on spectral lag, $N_{pk} \propto \Delta t^{-1}$,
will steepen to $N_{pk} \propto \Delta t^{-3}$.  By fitting the
$\Delta t^{-3}$ curve to intersect with GRB980425 one obtains a
complete lag-luminosity curve for GRBs which is consistent with
observed data (Figure \ref{lag-lum-980425}).  Knowledge of the shape
of this curve allows determinations of some key quantities of interest
for GRBs; particularly jet opening angle, maximum lorentz factor, and
total energy.

\section{ A Break in the Lag-Luminosity Relation }

In \citet{nmb00} was presented a relationship between the peak
luminosity of gamma-ray bursts (GRB) and the pulse time lag between
BATSE energy channels.  In \citet{jay00} this correlation was found to
be substantially improved when we neglected our poor knowledge of
received photon energy, thus taking the correlation between photon
number luminosity and pulse time lag.  It was found that the inferred
isotropic peak number luminosity $N_{pk}$ (photons sec$^{-1}$) is
related to the observed spectral lag between energy channels $\Delta
t$ by
\begin{equation}
N_{pk} = 8.6 \times 10^{56} \Delta t^{-0.98} ~.
\end{equation}

In \citet{jay00} was presented a kinematic interpretation of the
origin of this relation.  Specifically, bursts with emitting material
moving with a higher velocity toward the observer appear more luminous
and have shorter observed lag (derived from an intrinsic pulse cooling
timescale) between observed energy channels due to relativistic blue
shift.  Relativistic beaming allows one to only consider emitters
moving directly toward the observer.  I proposed that the wide range
of observed (cosmological redshift compensated) spectral lags and
inferred luminosities (see Figure \ref{lag-lum-980425}) could be
explained if GRBs derive from a relativistic jet in which the fastest
material moves along the core of the jet and the velocity of the
material monotonically decreases with increasing angle from the jet
axis.  The variety of observed bursts then derives from our
perspective of the jet.  All of the material is assumed to move
relativistically ($\gamma \gg 1$) and so all of our received flux is
derived from a very small $\sim 1/\gamma^2$ solid angle of the jet;
much smaller than the jet opening angle ($1/\gamma \ll \theta_0$).  It
is from this small region that all of our information about a burst is
derived.

The question then arises: what happens when the jet is observed at
such large angles from its central axis that the observed emitting
material is moving at `slow' enough velocity that the $1/\gamma$
beaming angle becomes comparable to the diametric angular size of the
emitter ($\theta_0 \sim 2/\gamma$)?  At this point the size and extent
of the emitting region becomes directly observable.  In the
relativistic case ($\gamma \gg 1$), the received flux emits from an
{\em apparent} region of size $R_s/\gamma$ where $R_s$ is the emitter
distance from the center of the source.  If $\gamma$ is small such
that $1/\gamma > \theta_0 \equiv R_e/R_s$, where $R_e$ is the size of
the emitting region, then the received flux emits from the entire
physical size of this region $R_e$, i.e. the physical extent of the
jet becomes observable.

Using this distinction one can derive the dependence of the inferred
isotropic number luminosity $N \propto f_N R^2$ on Lorentz factor for
the two cases, where number intensity $f_N = \gamma^3 f'_N$. The
relativistic $N_{rel}$ scales as
\begin{equation}
N_{rel} \propto (\gamma^3 f'_N) (R_s/\gamma)^2 \propto \gamma \quad \text{for $1/\gamma < \theta_0/2$}
\end{equation}
and the sub-relativistic $N_{sub-rel}$ scales as
\begin{equation}
N_{sub-rel} \propto (\gamma^3 f'_N) R_e^2 \propto \gamma^3 \quad
\text{for $1/\gamma > \theta_0/2$} ~.
\end{equation}
Thus, taking the observed spectral lag dependence $\Delta t_{obs}
\propto 1/\gamma$ \citep{jay00} one gets
\begin{equation}
N_{rel} \propto \Delta t^{-1} \quad \text{for $1/\gamma < \theta_0/2$} 
\end{equation}
\begin{equation}
N_{sub-rel} \propto \Delta t^{-3} \quad \text{for $1/\gamma > \theta_0/2$} ~.
\end{equation}
Thus if GRBs are jets, one can expect a break in the lag-luminosity
relationship.  This is analogous to the expected break in the
lightcurve of a GRB afterglow as it transitions from relativistic to
sub-relativistic expansion \citep{rhoads97}.

The inclusion of GRB980425 into the set of all long GRBs is the
simplest explanation; one need not invoke a separate phenomenon to
explain a burst so vastly weaker than its cosmological counterparts.
The model presented here unifies these seemingly disparate events by
way of a single mechanism; observer perspective on a relativistic jet.
For GRB980425 to be a member of the set of known bursts, it must be in
the sub-relativistic regime.  Thus I fit the $N_{sub-rel}(\Delta t)$
curve to contain GRB980425.  The result is a complete lag-luminosity
relationship for long GRBs and is shown in Figure
\ref{lag-lum-980425}.  Although the data is as yet very sparse, it is
consistent with the curve.

The intersection of the $N_{rel}(\Delta t_{obs})$ and
$N_{sub-rel}(\Delta t_{obs})$ curves is at
\begin{equation}
N_{int} = 5.5 \times 10^{57}\  \text{photons sec$^{-1}$}
\end{equation}
\begin{equation}
\Delta t_{int} = 0.15\ \text{sec} ~.
\end{equation}
At this point the jet and beaming angles are the same, $\theta_0
= 2/\gamma$.  As in \citet{jay00}
\begin{equation}
\frac{\gamma_{int}}{\gamma_{980425}} = \frac{\Delta t_{980425}}{\Delta
t_{int}} = \frac{4.5\ \text{sec}}{0.15\ \text{sec}} = 30
\end{equation}
so $\gamma_{int} = 30 \gamma_{980425}$.  Similarly for the brightest
burst in this dataset, $\Delta t_{990123} \simeq 0.01$, so
$\gamma_{990123} \simeq 450 \gamma_{980425}$.  The value of
$\gamma_{980425}$ is uncertain.  Bounds will be discussed in the next
section, but for now we take $\gamma_3 \equiv \gamma_{980425}/3$.
Thus the brightest, fastest bursts such as GRB990123 have Lorentz
factors $\gamma_{max} \sim 1000 \gamma_3$ or higher, while more
middling bursts such as GRB980703 have $\gamma \sim 100 \gamma_3$.

The diametric opening angle of the jet is
\begin{equation}
\theta_0 \equiv \frac{2}{\gamma_{int}} = \frac{2}{90 \gamma_3} \simeq
\frac{1.3^o}{\gamma_3} ~.
\end{equation}
In \citet{jay00} it was estimated that $\theta_0 \sim (5^o\ \text{to}\
10^o)/\gamma_{100}$ where $\gamma_{100} \equiv \gamma_{max}/100$,
which is in agreement with the value presently calculated.

Having knowledge of the opening angle of the jet, $\theta_0 \simeq
1/45/\gamma_3$, we may calculate the beaming factor 
\begin{equation}
\begin{split}
f_\Omega &= 1 - \cos\biggl(\text{Arcsin}\biggl(\frac{1}{\gamma} \biggr)
+ \frac{\theta_0}{2}\biggr)\\ &\simeq \onehalf \biggl( \frac{\theta_0}{2}
\biggr)^2 = 6.4 \gamma_3^{-2} \times 10^{-5}
\label{beamingfraction}
\end{split}
\end{equation}
where $1/\gamma \ll \theta_0/2 \ll 1$ for cosmological GRBs.  The
total energy in $\gamma$-rays is related to the inferred isotropic
energy by $E_{tot} = f_\Omega E_{iso}$.  This is a large reduction in
the energy required to make a GRB.  For instance, GRB990123 had
$E_{iso} = 2 \times 10^{54}$ ergs \citep{gbwv+99} and thus $E_{tot} =
1.2 \gamma_3^{-2} \times 10^{50}$ ergs.  So $\gamma$-ray energies of
bursts are a fraction of the total collapsar energy.

\section{ Event Rate Estimation }

Having derived the above quantities, one can now make some crude
estimations of rates of GRBs and SNe Ib/c.  First, what fraction of
SNe Ib/c make GRBs?  \citet{wmf99} estimate that 1\% of SNe Ib/c have
large enough helium cores to make a collapsar SN-GRB.  \citet{bkhp+98}
estimate the SN Ib/c rate to be 0.3/day out to distance of $100
h^{-1}_{65}$ Mpc, which is roughly the limiting distance for a
GRB980425 to be seen by BATSE.  Thus one expects 0.003/day $\sim
1$/year GRB rate within this distance.  \citet[see their
Fig. 2]{nbw99} find that the number of bursts similar to GRB980425 and
with comparable lags comes to roughly $\sim 1$/year, consistent with
this estimate.

A lower bound for our rate can give information on the Lorentz factor
of $\gamma_{980425}$.  The beaming factor $f_\Omega = 1 -
\cos(\text{Arcsin}(1/\gamma) + \theta_0/2)$ gives the probability of
observing an event's jet.  Since $1/\gamma \gg \theta_0$ for
GRB980425, $f_\Omega \approx 1 - \sqrt{1 - 1/\gamma^2} = 1 - v/c \sim
1/2/\gamma^2$ where $v$ is the speed.  Since we estimate a SN-GRB rate
of 1/yr within 100 Mpc, the inverse average number of years between
detected SN-GRBs estimates $f_\Omega$ which, in turn, gives a weak
(due to incompleteness of SN-GRB detections) lower bound on $\gamma$.
For instance, if $\gamma = 5$, then $f_\Omega = 0.02$, and we can
expect a burst like GRB980425 every $\sim$ 50 years. For $\gamma = 2$
and 3 one gets 7 and 17 years respectively.  \citet{bkhp+98} and
\citet{nbw99} find little evidence for another convincing SN-GRB event
in the BATSE catalog.  Firstly, this is not a problem; GRB980425-like
events are likely rare enough to have been missed or not exist during
BATSE mission duration. Secondly, assuming completeness (a poor
assumption) the data begins to push the lower bound on $\gamma$ up to
around 2 or 3.  \citet{kfwe+98} argue for a radio emission source
moving with $\gamma \sim 2$, while \citet{wmf99} suggest that $\gamma
\approx 5$ could have created GRB980425.  These values likely bracket
the real value.  Thus in this paper I assume $\gamma_{980425} \sim 3$.

At cosmological distances, taking the above rate of GRB generating
SNe, one gets a rate out to radius $r$
\begin{equation}
\begin{split}
\text{Rate}(r) &\sim f_\Omega\ 3 \times 10^{-3}/\text{day}\
\biggl(\frac{r}{100 \text{Mpc}} \biggr)^3\\ &= 0.2/\text{day} \biggl(\frac{r}{10
\text{Gpc}} \biggr)^3 ~.
\end{split}
\end{equation}
using Eqn (\ref{beamingfraction}).  This is consistent with the $\sim$
1 GRB/day observed by BATSE.  Much better modeling including cosmology
and star formation rates can be done.  However, the point to make here
is that it is consistent that GRB980425 and the cosmological bursts
derive from the same progenitor.

\section{ Discussion } \label{sectionbeaming}

The obvious prediction of the model present herein is that all (long)
GRBs will fall along the curve in Figure \ref{lag-lum-980425}.  In
addition, since this model supports the collapsar model for GRB
progenitors, all (long) GRBs should have a SN buried within as
predicted by \citet{wmf99}.

Many questions remain.  For instance GRB980425 had a quickly decaying
X-ray afterglow \citep{bkhp+98}.  Perhaps numerical simulations of
afterglows from relativistic jets \citep[e.g.][]{gmps00} could yield
insights into this behavior.  If this is a common characteristic of
off-axis jets, this might have an effect on the predicted rates of
observable so called ``orphan afterglows'' \citep{rhoads97}.

This work was performed under the auspices of the U.S. Department of
Energy by University of California Lawrence Livermore National
Laboratory under contract W-7405-ENG-48.

\clearpage

\end{document}